\documentclass{epl}
\usepackage{amsbsy}
\usepackage{amssymb}
\usepackage{subfigure}

\title{Ageing, dynamical heterogeneities and crystallization in the Biroli-Mezard model}
\shorttitle{Dynamical behaviour of the BM model}
\author{E. Marinari\inst{1} \and V. Van Kerrebroeck\inst{1}}

\institute{                    
  \inst{1} Dipartimento di Fisica, Universit\`a di  Roma {\em La Sapienza} - P.le Aldo Moro 2, 00185 Roma, Italy\\
}
\pacs{66.43.Fs}{Glasses}
\pacs{61.20.Lc}{Time-dependent properties of liquid structure; relaxation}
\pacs{05.50.+q}{Lattice theory and statistics}

\begin{document}

\maketitle

\begin{abstract}
We discuss a number of essential dynamical features of the
Biroli-Mezard model.  We observe a dynamical slowing down which for
high densities depends both on the observation time and on the age of
the system. We relate this ageing behavior to the real space dynamics
of the system, from which it is clear that the system initially gains
mobility due to some essential initial restructuring.  We discuss
crystallization, its relevant time scales and connection with ageing.
We quantify how heterogeneous the slow dynamics of the BM model is by
investigating the dynamical susceptibility and discussing its
dependence on the age of the system.  We establish a number of
quantitative evidences that are in good agreement with a recent
analysis.
\end{abstract}

\section{Introduction}
During the last several decades many features of glassy systems have
been studied in remarkable detail. New, accurate experiments have
allowed to establish properties such as dynamical slowing down, the
presence of dynamical heterogeneities, ageing and other memory effects
which are typical of glassy systems~\cite{cugl02}. Nonetheless, a true
theoretical understanding is lacking~\cite{ediger}.  There are
theories which see the glass transition as a kinetically disguised
thermodynamic singularity, such as the entropic theory and the free
volume theory. Others, \emph{e.g.}~mode coupling theory, interpret 
the transition as purely dynamic.  
Kinetically constrained models~\cite{kcm}
were initially introduced to assess the extent to which glassy
behavior could be understood, while assuming only trivial equilibrium
behavior.  In particular, lattice gas models with their fixed
particle density describe mainly structural glasses. They are
hard-sphere models, such that each lattice site can be occupied by at
most one particle.

There are two main classes of lattice gas models.  A first class
induces glassy behavior by introducing dynamical rules governing the
conditions under which a particle can move to a nearest neighbor
site. Since there are no restrictions on the allowed configurations,
these lattice gasses obviously have a trivial equilibrium behavior and
a flat energy landscape. 
The Kob-Anderson~\cite{ka,fra-mul-par,ton-bir-fish,mar-pit} 
model is probably
the most archetypal example of these types of lattice gasses.  A
second class of lattice gas models is represented by the
Biroli-M\'{e}zard (BM) model\cite{bm}, defined by introducing a
geometric constraint. More precisely, BM prescribes that each particle
can have at most $l$ neighbors. The essential difference with the
former is that the BM model is defined through a thermodynamic
constraint, which only allows certain types of configurations.  
The KA-model, on the other hand, places no restrictions on which
configurations are allowed; only its dynamics is governed by
the kinetic constraint.

In this paper we discuss the ageing dynamics in the BM model. We are mainly involved
with the identification of different types of relaxation dynamics in different time windows.
We discuss crystallization in the model and analyze the $4$-point susceptibility and its 
large time and large volume behavior. 

\section{Model Specifications}

For our realization of the BM model we use the same type of model and
preparation method proposed in the original work by Biroli and 
M\'ezard~\cite{bm}. 
In order to avoid crystallization~\cite{fin-en-BM}, we consider a 
three-dimensional mixture of $30\%$ of particles characterized by 
$l=1$ and $70\%$ of particles with $l=3$.
The initial configurations are obtained by performing an annealing 
procedure on a random assignment of $N$ particles until we find a 
valid configuration.  
The considered densities are smaller, but close to the value 
$\rho_{\mbox{\tiny{c}}}\simeq 0.565$ found in~\cite{bm} 
which has been hypothesized to be the
density around which a dynamical phase transition takes place and
hence, beyond which there is a full dynamical arrest.
The reported results are obtained from simulations on cubical lattices of 
linear size $L$ ranging from $10$ to $30$. While simulations of systems of 
size $L=10$ are computationally less demanding, they start equilibrating 
faster (\emph{i.e.}, during our accessible running time). 
So, for a more 
qualitative analysis or when focusing on the initial time-dependent 
behavior, we consider smaller samples. Conversely, when characterizing the 
long-time slow dynamics, we turn to the larger systems, which do not show 
any signs of equilibration.

\section{Ageing Slow Dynamics}

\begin{figure}
\centering
\input{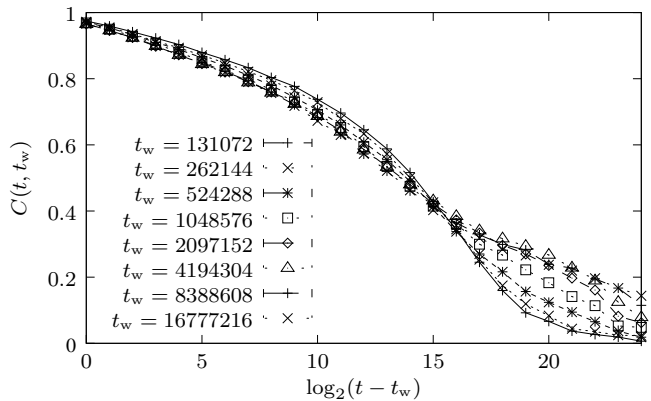}
\caption{Two-time correlation functions $C(t,t_{\mbox{\tiny{w}}})$ as a function of
the logarithm of the elapsed time $t-t_{\mbox{\tiny{w}}}$ (since the waiting time
$t_{\mbox{\tiny{w}}}$), for a system of linear size $L=30$ and particle density
$\rho=0.542$. During the initial time window, the upper curves correspond to shorter
waiting times, while at later elapsed times they are the lower ones.}
\label{fig:ttcorr}
\end{figure}

In order to study the dynamics, and more specifically the ageing
behavior of the BM model we investigate the two-time correlation
function~\cite{biroli} 
\begin{equation}
\label{eq:bm-ttcorr}
C(t,t_{\mbox{\tiny{w}}}) = \frac{1}{V\rho(1-\rho)} 
\sum_{i}\left[n_i(t)n_i(t_{\mbox{\tiny{w}}})-\rho^{2}\right]\;.
\end{equation}
Here $V$ is the volume of the system, \emph{i.e.}, the total number of sites,
and $n_i$ is the occupation number of site $i$, either at the waiting 
time $t_{\mbox{\tiny{w}}}$ (the time elapsed since the preparation of the system), or 
at time $t$, \emph{i.e.}, the actual age of the system (the total simulation time). 
Thus, the two-time correlation function measures the probability of 
finding site $i$ occupied at time $t$, if that same site was occupied 
at the waiting time $t_{\mbox{\tiny{w}}}$. 
In fig.~\ref{fig:ttcorr} we show our results for the largest and 
most dense system that we have investigated.  
Clearly the system relaxes in a non-exponential way.
Moreover, it ages, though it does not 
 fully resemble the typically expected behavior of ageing systems
for which short time effects usually do not depend on the age of the system, but
the long time dynamics does~\cite{biroli}.  Instead, we find that for the BM model
the dynamic correlation function depends at any time also on the actual age of the
system.  We distinguish an initial ``anti-ageing'' regime, during which
younger systems decay slower than their older counterparts, from a later more
standard type of ageing, where older systems relax slower.

As mentioned, the initial acceleration of the dynamics of older systems
is generally not observed in ageing systems. However, in the case of the 
BM model it is inherent to the way we create our valid BM configurations. 
The annealing procedure we apply to a less constrained structure could 
also be considered as a (soft) quench to zero temperature dynamics, 
which causes the initial dynamics to be somewhat atypical. 
When we will discuss the real space picture it will become clear that 
the observed acceleration is due to some restructuring which only takes 
place during an intermediate time window.  
In fact, we can define a time $t_{A}^L(\rho,t_{\mbox{\tiny{w}}})$ where the initial 
acceleration is overcome by standard ageing. As one would expect $t_{A}$ 
increases with density and system size. It also reveals to be slightly 
dependent on $t_{\mbox{\tiny{w}}}$, reassuring us that $t_A$ is not a clear-cut separation of two 
distinctive types of dynamics, but rather is a useful indicator at 
which the initial acceleration is overcome by standard ageing behavior.

Clearly, it is the time window starting from $t_{A}$ which is of
particular interest. Though all systems of any age will eventually
completely decorrelate, we find that older systems do so more
slowly. The same ageing effect has already been observed for some
kinetically constrained models which can be described by coarsening
dynamics~\cite{kcm}. In the case of lattice gas models, such
as the KA model, ageing can be observed by adding an external field
which allows for quenching or annealing of the 
density~\cite{ageing-ka,sellitto}. 
How to interpret the introduction of 
such an external field is not so straightforward, however.  
In the case of the BM model, the quench is not in terms of some external
field, rather it is a real quench with respect to the energy landscape. The
thermodynamic constraints are such that the energy landscape is
partitioned into configurations of either zero or infinite
energy. Though, it certainly is a more simplistic energy landscape
than that of real glasses, for which a valley may contain many
sub-valleys, it does have the essential quality that if two
configurations are separated by long pathways in configuration space
it will take a lot of local rearranging for the system to be able to
switch between them.  
When we have just created a configuration, little is known about its surrounding
energy landscape. The system starts exploring it, meaning it moves 
away from and forgets about the initial configuration. 
Thus, the fact that older systems eventually decorrelate more slowly suggests 
the system explores more isolated regions of phase space at later times.

We have tried to classify the ageing dynamics according to its behavior
as a function of $t$ and $t_{\mbox{\tiny{w}}}$. However, in our case, the
initial acceleration effect does not allow us to determine some
specific scaling law in a rigorous way.

Eventually, for less dense systems of linear size $L=10$ we observe the system
decorrelates completely. 
However, denser and larger ageing systems do not completely decorrelate during our running time.
\section{Real Space Structure}

\begin{figure}
  \centering
  \subfigure{\input{dynsk10_535-newa.pstex_t}\label{fig:struct-fact}}
  \subfigure{\includegraphics[width=5cm, height=5cm, angle=0]{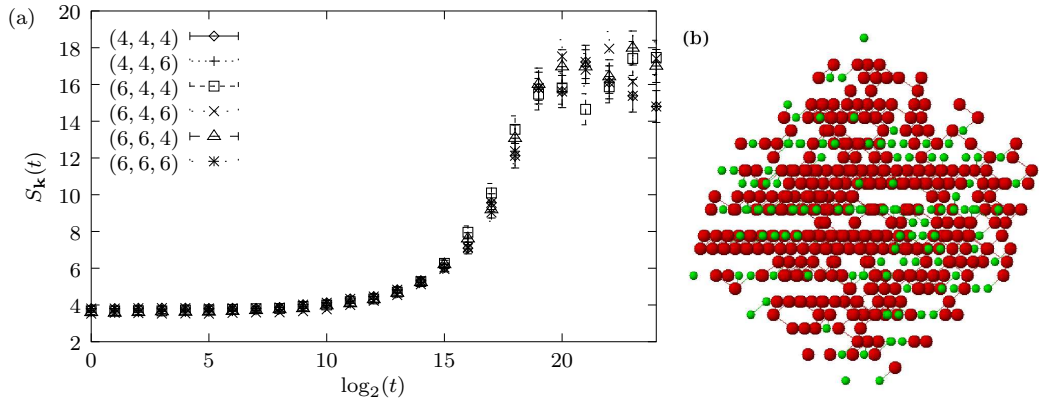}\label{fig:visL10_end}}
  \caption{(a) Dynamic structure factor and (b) stratification (at the end of the simulation) of a BM configuration of linear size $L=10$ with density $\rho=0.535$.}
\end{figure}

Motivated by the belief that a clearer view on the real space
structure of the system can help us reach a better understanding of
the (slow) dynamics, 
the ageing behavior, and in particular of the
intermediate acceleration of the dynamics of the BM model, we compute the
dynamic structure factor
\begin{displaymath}
S_{\bf k}(t)\equiv \frac{\langle \rho_{\bf k}^* \rho_{\bf k} \rangle}{N}\;,
\end{displaymath}
where $\rho_{\bf k}\equiv\sum_{i=1}^{N}\exp(i{\bf k}\cdot{\bf r}_i)$.
While in most directions we observe that $S_{\bf k}(t)$ does not grow
at all during our simulation time, we find that at intermediate times it does
increase along some of them. 
In fig.~\ref{fig:struct-fact} we show $S_{\bf k}(t)$ as a function of the total simulation time $\log_2(t)$ along these directions for the smaller systems. Larger systems reveal the same type of behavior, though at a slightly later time.
This figure clearly shows that the growth of the
structure factor is restricted to an intermediate time interval, after 
which $S_{\bf k}(t)$ remains constant.
Even though the final value of the structure factor is
small, it indicates that our system is subject to a modest degree of
crystallization or at least restructuring.  In spite of this effect,
from our analysis of the correlation function it turns out that at
later times the system does have a glassy behavior. 
In fact, in~\cite{cav-giar} it was pointed out that there exist models in which it is not 
necessarily easy to distinguish the dynamics of a highly disordered polycrystal 
or a locally stable glassy phase. Thus, the finding that the system 
reveals some tendency to crystallize, does not necessarily need interfere with the 
observation of some typical glassy features.

In order to clarify what type of restructuring is occurring, we
visualize an exemplar BM system at some late instance of the
simulation in fig.~\ref{fig:visL10_end}. The picture clearly shows
that the system has reorganized itself into a stratified
structure along a direction close to the body diagonal of the cubical
sample we considered, in agreement with the directions along which the
structure factor grows. Both samples of linear size $L=10$ and $L=30$
reveal this alternation of several more densely populated planes with
usually one that is less occupied. 
However, given the irregularity of
the alternation, it does not seem that the observed stratification is
the beginning of an overall crystallization of the system, nor does a
closer investigation with the naked eye allow to identify any
recurring structure along these planes.
\begin{figure}
  \centering
  \input{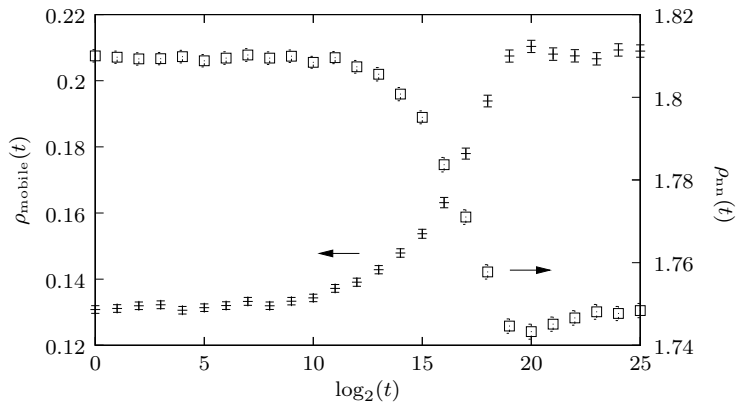}
  \caption{Average number of neighbors of particles ($\square$) and average number of mobile particles ($+$) in function of the total running time for a system of linear size $L=10$ with density $\rho=0.535$.}
\label{fig:non-mob}
\end{figure}

Even though the system does not 
go through a complete crystallization, it is
important to consider what consequences this restructuring
has on the systems dynamics. Therefore, we compute the average number of neighbors of each
particle $\rho_{\mbox{\tiny{nn}}}(t)$ and the overall density of mobile particles $\rho_{\mbox{\tiny{mobile}}}(t)$, \emph{i.e.}, particles
which are allowed to move to one of the neighboring empty sites,
as a function of time $t$. We present these
quantities
in fig.~\ref{fig:non-mob},
for the smaller system with $L=10$. We see that on average the
number of neighbors of a particle undergoes a considerable decrease
during  the same time interval where the dynamic structure factor
grows. 
Also in that same time region, the  density of mobile particles $\rho_{\mbox{\tiny{mobile}}}$ increases. 
After this, both quantities still fluctuate, but
basically the system maintains its newly acquired mobility. Thus, the
stratification results in a structure where, on average, particles
have less neighbors. As a consequence, they are less constrained, hence more of them become mobile.
It is important to note that the larger amount of mobile particles
is not just a simple reflection of the resulting reduced $3D$ layered
structures. Though they originate from the restructuring of the
system, they are a characteristic of the whole system.

This allows us to explain what is really happening during the initially observed
acceleration of older systems described in the previous section. We have just
shown that because the system is restructuring it gains in
mobility. Thus, in general, older systems will be characterized by a
higher overall mobility for short elapsed times $t-t_{\mbox{\tiny{w}}}$ and initially
they will decay faster than younger, less mobile systems. However,
once this initial time interval has been overcome, \emph{i.e.}, the system
has restructured in order to become mobile, older systems decorrelate
slower than younger ones.
Hence, there must be some explanation, other than this 
stratification, as  to why we observe the slower decorrelation of older 
systems at larger times, as discussed in the previous section.

\section{Intrinsically Heterogeneous Dynamics}

A quantity which typically reveals the degree to which glasses are
dynamically heterogeneous, is the $4$-point susceptibility, defined as
the variance of the correlation function,
\begin{equation}
\label{eq:bm-chi}
  \chi_4(t,t_{\mbox{\tiny{w}}})\equiv 
  V \times \left( \langle C(t,t_{\mbox{\tiny{w}}})^2 \rangle 
  - \langle C(t,t_{\mbox{\tiny{w}}}) \rangle^2 \right)\;.
\end{equation}
A rigorous analysis of the theoretically expected behavior for
$\chi_4(t)$ is given in~\cite{ton-bouch}. In full generality, it is
possible to distinguish several time regimes during which the dynamic
susceptibility usually follows a power-law, $\chi_4(t) \sim
t^\mu$. Here, the actual value of the parameter $\mu$ depends strongly
on the dominating relaxation mechanism. We shall compare our results
with these predictions and extend the discussion to the dependency on
the waiting time, for which --to our knowledge-- there are no such
qualitative predictions available yet.
\begin{figure}
\centering
\input{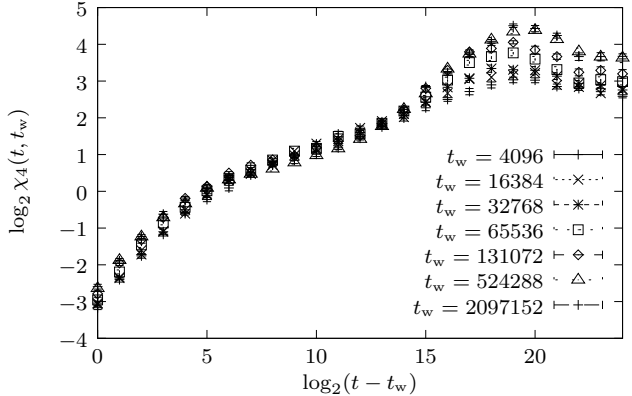}
\caption{Four-point dynamic susceptibility $\chi(t,t_{\mbox{\tiny{w}}})$ in function
of the logarithm of the elapsed time $t-t_{\mbox{\tiny{w}}}$ since the waiting time 
$t_{\mbox{\tiny{w}}}$, for systems of linear size $L=10$ with density $\rho=0.535$. 
At large elapsed times, the curves correspond to systems of 
increasing ages from bottom to top.}
\label{fig:chi4}
\end{figure}

We present the two-time temporal behavior for systems of linear size
$L=10$ and density $\rho=0.535$ in fig.~\ref{fig:chi4}.  During a
small initial (elapsed) time window, the dynamic susceptibility grows
according to a power law with $\mu \simeq 0.6$. This is in 
reasonable agreement with the $\mu=0.5$ prediction by~\cite{ton-bouch}
for the elastic regime. For such an elastic regime we expect no
dependency on the age of the system, as is confirmed by our results.

The subsequent time regime is marked again by a power-law, but now
with $\mu\sim 0.2$.  This matches the early $\beta$-regime and the
subsequent plateau described by mode-coupling theory (MCT) for the
correlation function. Although, we do not have any prediction for the
MCT-parameters for the BM model, the local structural relaxation
associated to the $\beta$-regime explains the observed behavior.  
Also, it is in this time window that the acceleration of older systems was 
observed, though it is no longer visible for the dynamic susceptibity.
In fact, if it is just due to the way in which we prepare our BM system,
it should not be strongly sample dependent and therefore not
contribute to the variation of the correlation function.

Conversely, the next regime is distinguishable by the fact that the
susceptibility does become dependent on both $t$ and $t_{\mbox{\tiny{w}}}$.  In fact,
this regime sets in at a time of the order of $t_{A}(t_{\mbox{\tiny{w}}},\rho)$. In
MCT we can map this with the late $\beta$-regime which continues into
the so-called $\alpha$-regime. In fact, one expects the system to
become increasingly two-time dependent when the system relaxes in an
increasingly collective manner. From the correlation function it was
clear that older systems relax a lot slower than their younger
counterparts. If the slow dynamics is cooperative in nature, we expect
older systems to have an increasingly higher value for the dynamic
susceptibility. Indeed, fig.~\ref{fig:chi4} confirms that older
systems are increasingly more dynamically heterogeneous during this
regime.

The three time windows described up to now exist and are well defined 
for all systems that we have investigated. 
The remainder of the discussion, however, is observed 
only for smaller, less dense systems and hence those which completely loose their memory 
during our running time. Larger and denser systems do not decorrelate 
completely during our simulations, but neither does the dynamic
susceptibility peak during our running time. Rather, for the latter 
$\chi_4(t)$ is only observed to grow.

For the system represented in fig.~\ref{fig:chi4} the growth of
$\chi_4(t,t_{\mbox{\tiny{w}}})$ continues up to some time $t^*-t_{\mbox{\tiny{w}}}$ for systems of any
age. Though we cannot say with certainty that the time at which
$\chi_4(t)$ peaks, \emph{i.e.}, for $t^*-t_{\mbox{\tiny{w}}} \sim 2^{19}$, is independent of
the age of the system, it seems to be less influenced than the value 
of $\chi_4$ at the peak itself. 
Being the time at which we observe the system has completely decorrelated 
and given the log-scale this is not so surprising though.  
In line with expectation of the preceding time window, the
peak of $\chi_4$ is larger for older systems indicating that the
latter are more dynamically heterogeneous after shorter elapsed times.

After having peaked, the susceptibility decreases and reaches a
constant value when the system has thermalized. 
Although~\cite{ton-bouch} suggested the subsequent decay
should be very fast, it is not clear to us why this should necessarily
be the case. At least for BM we find the subsequent decay to follow
again a power-law, characterized by a largely $t_{\mbox{\tiny{w}}}$-independent value
for $\mu \sim 0.2$. This indicates the system is probably already
closer to its steady state.
In fact, the final value of the
susceptibility, reached within our running time by the younger
systems, is lower than the peak-value, but higher than one, the value
predicted for lattice gases with trivial equilibrium~\cite{kcm}.  The
limit of the susceptibility for the BM model, which has non trivial
equilibrium configurations, is not so easily calculated. However, from
our simulations, we find that it reaches a constant value, which is
density and waiting time dependent.
The exact dependence on the density of this limiting
value is not clear, though it seems to grow with density.  In any
case, it suggests that the BM system is inherently
dynamically heterogeneous. This is also supported by the fact that 
the final value of the susceptibility of older systems is larger than that of younger ones.

\section{Conclusions}
In this paper, we have analyzed the dynamical behavior of the
thermodynamically defined BM lattice glas model.  We have found some
interesting ageing behavior, and we have discussed its connections
with the non trivial topology of the energy landscape of these models.
We have analyzed the real space behavior of the model and found signs
of a layered structure emerging with time. The reorganized structure
is however not reflected in the mobility of the particles.  The
dynamics is observed to be increasingly heterogeneous for older
systems.  Moreover, when the system has lost its memory of the initial
configuration, the $4$-point-susceptibility assumes a
density-dependent constant value suggesting an important role of
intrinsic heterogeneities in the model.  A comparison with predictions
of \cite{ton-bouch} is very successful for a large number of features.
It only breakes down regarding the rate of decay after the
susceptibility peaks: this is one of the points that obviously calls
for further clarification.

\acknowledgments 
We thank Andrea Cavagna, Kenneth Dawson, Irene Giardina, Walter Kob,
Estelle Pitard and Francesco Sciortino for interesting conversations.
This work was supported by EVERGROW, integrated project No.~1935 in
the complex systems initiative of the Future and Emerging Tecnologies
directorate of the IST Priority, EU Sixth Framework.

\end{document}